\documentclass[showpacs,twocolumn,amssymb]{revtex4}
\usepackage{graphicx}
\usepackage{dcolumn}
\usepackage{bm}
\usepackage{epsfig}
\usepackage{amsmath}
\usepackage{ulem}
\newcommand{\ud}{\mathrm{d}}
\begin{document}

\title{Precise peculiar velocities from gravitational waves accompanied by electromagnetic signals and cosmological applications}
\author{Y. Y. Wang$^1$, F. Y. Wang$^{1,2\ast}$ and Y. C. Zou$^{3\dag}$}
\affiliation{$^{1}$School of Astronomy and Space Science, Nanjing
University, Nanjing 210093, China \\
$^{2}$Key Laboratory of Modern Astronomy and Astrophysics (Nanjing
University), Ministry of Education, Nanjing 210093, China \\
$^3$ School of Physics, Huazhong University of Science and Technology, Wuhan 430074, China\\
$^\ast$Electronic address: fayinwang@nju.edu.cn\\
$^\dag$Electronic address: zouyc@hust.edu.cn}

\date{\today}

\begin{abstract}
{Peculiar velocities are a precious tool to study the large-scale
distribution of matter in the local universe and test cosmological
models. However, present measurements of peculiar velocities are
based on empirical distance indicators, which introduce large error
bars. Here we present a new method to measure the peculiar
velocities, by directly estimating luminosity distances through
waveform signals from inspiralling compact binaries and measuring
redshifts from electromagnetic (EM) counterparts. In the future,
with the distance uncertainty of GW events reducing to $0.1$ per
cent by future GW detectors, the uncertainty of the peculiar
velocity can be reduced to $10$ km/s at 100 mega parsecs. We find
that dozens of GW events with EM counterparts can provide a Hubble
constant $H_0$ uncertainty of $0.5\%$ and the growth rate of
structure with a $0.6\%$ precision in the third-generation
ground-base GW detectors, which can reconcile the $H_0$ tension and
determine the origins for cosmic accelerated expansion.}
\end{abstract}


\maketitle

\section{Introduction}\label{sec:intro}
Recently, the Advanced LIGO and Advanced Virgo discovered the first
gravitational wave (GW) signal from coalescing binary neutron stars
accompanied by electromagnetic (EM)
counterparts\cite{Abbott17a,Abbott17c,Goldstein17}. This breakthrough heralds
the new era of gravitational-wave multi-messenger astronomy. GW
measurements of coalescing binaries can make cosmological
measurements. Schutz first pointed out that the waveform signal
from inspiralling compact binaries can be used to measure the
luminosity distance to the source with high
precision\cite{Schutz86}. GW standard sirens can probe the cosmic
expansion history and the dark energy with high
accuracy\cite{Holz05,Cutler09}. Similar to standard candles, an
independent measure of the redshifts of EM counterparts is crucial.
Mergers of binary neutron stars (BNSs) or neutron star-black hole
(NS-BH) binaries are the most promising GW sources accompanied by
detectable EM counterparts. The discovery of EM counterparts of
GW170817 has realized this idea\cite{Abbott17c}. Therefore, GWs
together with EM counterparts providing the redshift information,
could be an excellent cosmological probe.

When using GWs and EM counterparts to measure the distance-redshift
relation, the redshifts should be entirely due to the cosmic
expansion. However, in the local universe, large-scale structure
induces peculiar motions so that the measured redshifts contain
contributions from peculiar velocities\cite{Peebles93,Strauss95}.
Meanwhile, the horizon of Advanced LIGO and Advanced Virgo for the merger of
BNS or NS-BH is only a few hundreds Mpc\cite{Abbott16}. If we take
the typical value of peculiar velocity $v_{\text{pec}}=400$~km~s$^{-1}$, and
Hubble constant $H_0=70$~km~s$^{-1}$~Mpc$^{-1}$, the peculiar
velocity can contribute about 30\% of the measured redshift at a
distance of 20 Mpc. Therefore, the effect of peculiar velocities on
GW astronomy is crucial in local universe. Previous works using GW
and EM counterparts as cosmological tool do not consider this
effect. Meanwhile, peculiar velocities are important for directly
probing the distribution of dark matter\cite{Peebles93}, studying
precision cosmology from type Ia supernovae (SNe
Ia)\cite{Hui06,Gordon07}, measuring the Hubble
constant\cite{Ben-Dayan14}, probing the growth rate\cite{Guzzo08}
and redshift space distortion\cite{Kaiser87,Zhang13}. In
consequence, measurement of peculiar velocity is of great importance
for cosmology\cite{Huterer18}.

At present, there are two ways to measure peculiar velocities. The
first method is to measure peculiar velocities directly by obtaining
distances to individual galaxies and their redshifts. Therefore, the
accuracy of peculiar velocities depends on distance indicators. Many
distance indicators independent of Hubble constant have been used,
including SNe Ia\cite{Riess97}, Tully-Fisher relation\cite{Tully77,
Theureau07, Springob07} and the fundamental plane
relation\cite{Djorgovski87}. However, there are several sources of
systematic error, such as Malmquist bias, luminosity evolution, and
imperfect corrections for dust extinction for SNe Ia. At the same
time, the Tully-Fisher relation and the fundamental plane relation
yield individual distance uncertainties of
20\%-25\%\cite{Strauss95,Willick97}. The second method measures the
peculiar velocities statistically based on redshift space
distortion\cite{Kaiser87}. Unfortunately, the peculiar velocities
derived from both methods have large uncertainties\cite{Huterer18}.

Below, we calculate the peculiar velocity of the host galaxy of
GW170817. By fitting the waveform signal of the GW170817, the
distance of this event is 43.8$^{+2.9}_{-6.9}$ Mpc\cite{Abbott17b}.
From the EM counterparts, the heliocentric redshift of host galaxy
NGC 4993 is $z=0.009783$\cite{Levan17}, which corresponds to $z =
0.01083$ in the CMB frame. In order to derive the peculiar velocity
from equation (1), the Hubble constant should be known. Recent
measurement of the local Hubble constant from SNe Ia is claimed to
be accurate at the 2.4\% level\cite{Riess16}, suggesting
$H_0=73.24\pm 1.74 ~\rm km~s^{-1} Mpc^{-1}$. However, this value is
3.4$\sigma$ higher than $67.8\pm0.9~\rm km~s^{-1} Mpc^{-1}$
predicted by $\Lambda$CDM model from Planck cosmic microwave
background (CMB) data\cite{Planck16}. The radial peculiar velocity is
$v_{\text{pec,r}}\simeq 276~\rm km~s^{-1}$ if $H_0$ from Planck is used.
Using the error propagation formula, the uncertainty of
$v_{\text{pec,r}}$ is about 200 $\rm km~s^{-1}$. The radial peculiar velocity
turns out to be $v_{\text{pec,r}}\simeq 38~\rm km~s^{-1}$ for $H_0$ from
SNe Ia\cite{Riess16}. So, the radial peculiar velocity is heavily
dependent on the Hubble constant $H_0$. In local universe, they
degenerate with each other.

The peculiar velocity of NGC 4993 was also derived using a
dark-matter simulation from the Constrained Local Universe
Simulations project\cite{Hjorth17}. However, the initial conditions
of peculiar velocities are derived from the Tully-Fisher relation
and the fundamental plane relation\cite{Hjorth17}. By assuming a
small error of recession velocity, the peculiar velocity with a
small error is obtained. Abbott et al. (2017b)\cite{Abbott17b} used
the 6dF galaxy redshift survey peculiar velocity
map\cite{Springob14} to derive the peculiar velocity of NGC 4993.
The uncertainties of peculiar velocities depend on the errors of
distance and Hubble constant. Generally they are proportional to
$d_L$. We compare peculiar velocities' uncertainties for different
methods at a distance of 100 Mpc. With this distance, the errors of
peculiar velocities from SNe\cite{Riess97} are in the range of
$(500, 1000)$ $\rm km~s^{-1}$. From SFI++ data set\cite{Springob07}, they are
roughly $(800, 3000)$ $\rm km~s^{-1}$ at 100 Mpc from the galaxy
survey with Tully-Fisher relation. These errors together with that
of NGC 4993 (GW170817) are listed in figure 1 for comparison.

\section{Method and results}
Here we propose a robust method to measure the peculiar velocities
with high accuracy using GWs and EM counterparts. Based on the
observed redshift $z$ and the luminosity distance $d_L$ derived from
GW waveform signal, the radial peculiar velocity $v_{\text{p}}$ can be
calculated by
\begin{equation}
v_{\text{pec,r}} = cz - H_0 d_L,
\end{equation}
where $c$ is the speed of light and $H_0$ is the Hubble constant. Because of the
$H_0$ tension\cite{Freedman17}, we should provide an independently measured $H_0$ to calculate $v_{\text{pec,r}}$. If $H_0$
is measured precisely in the future, $v_{\text{pec,r}}$ can be derived directly from equation (1).

Our method is called $v-v$ comparison
method\cite{Riess97,Ma12,Hudson12}. It is performed by comparing the
radial peculiar velocities from GW observations with those reconstructed
from the galaxy survey.

\begin{figure}
  \includegraphics[width=0.45\textwidth]{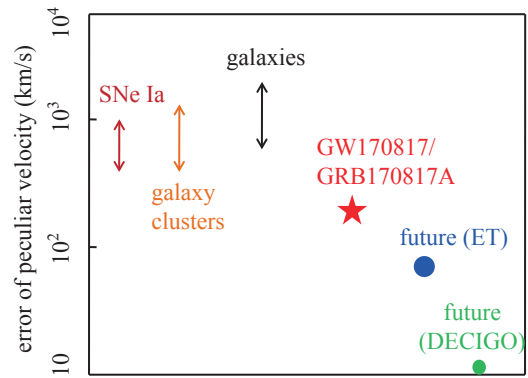}
  \caption{Measurement accuracy of the peculiar velocity at 100 Mpc for different methods.
  The method with label `SNe Ia' is from Riess et al. (1997) \cite{Riess97}, with label `galaxy clusters' is
  from Colless et al. (2001) \cite{Colless01}, and with label `galaxies' is from Springob et al. (2007) \cite{Springob07}.
  The error from our method is the star with label `GW170817/GRB170817A'. The future capable uncertainties of peculiar velocities
  are labelled as 'future (ET)` and `future (DECIGO)'. }
  \label{fig1}
\end{figure}

In the standard
$\Lambda$CDM cosmology, gravitational instability induces the growth
of density perturbations, which generate the peculiar velocity
field. In the regime where the density perturbation is linear, the
peculiar velocity ($\vec{v}_{\text{pec}}$) can be expressed
as\cite{Peebles93}
\begin{equation}
    \vec{v}_{\text{pec}}(\vec{x}) =
    \frac{H_{0}f_{0}}{4 \pi }\int \ud^{3}\vec{x}^{\prime
    }\delta _{\rm{m}}(\vec{x}^{\prime },t_0)\frac{(\vec{x}^{\prime }-\vec{x})}{%
    \left\vert \vec{x}^{\prime }-\vec{x}\right\vert ^{3}},
\end{equation}
where $f_{0}$ is the present day growth rate of structure, and $\delta_{\rm{m}}=(\rho-\overline{\rho})/\overline{\rho}$ is the
dimensionless density contrast. The growth rate of structure at scale factor $a$ is defined as
\begin{equation}
	f(a) \equiv \frac{\ud \ln D(a)}{\ud \ln a}
\end{equation}
where $D$ is the linear growth factor\cite{Peebles93}.

It is common to assume that a linear
bias exists between galaxy fluctuations $\delta_{\rm{g}}$ and matter
fluctuations $\delta_{\rm{m}}$ by introducing the bias parameter
$b$, i.e., $\delta_{\rm{g}} = b \delta_{\rm{m}}$. So the growth rate
of density fluctuations $f$ can be replaced with the parameter
$\beta \equiv f/b$. The $\beta$ can be combined with
$\sigma_{8,\text{g}}$ to get the growth rate since $f\sigma_8 =
\beta\sigma_{8,\text{g}}$.

Branchini et al. (1999) presents a self-consistent non-parametric
model of the local peculiar velocity field derived from the
distribution of IRAS galaxies in the Point Source Catalogue redshift
(PSC$z$) survey\cite{Branchini99}. The catalogue contains 15795
galaxies and the peculiar velocity field is reconstructed assuming
$\beta = 1$. The true peculiar velocities are proportional to $\beta$. Since the value of $\beta$ is uncertain, we need to constrain $\beta$ and $H_0$ simultaneously by comparing the peculiar velocities from equation (3) with measured peculiar velocities. Galaxies in the PSC$z$ redshift catalogue were used to
trace the underlying mass density field within 300 $h^{-1}$ Mpc
under the assumption of linear and deterministic
bias\cite{Radburn-Smith04}. Using the iterative
technique\cite{Yahil91}, the model velocity field is obtained from
the positions of galaxies in redshift space according to equation
(2). In our calculation, we interpolate model velocities at the
positions of mock GW events to compare predicted and observed
velocities. The predicted velocities in GW location are calculated
by applying a Gaussian kernel of radius $R_j$ (5 Mpc in this paper)
to the predicted 3D velocity $\textbf{v}_{\text{rec}}(\textbf{x}_j)$
specified at the position of the PSC$z$ galaxies. We have
\begin{equation}
    \textbf{v}_{\text{smo}}(\textbf{x}_i) = \frac{\sum_{j=1}^{N'}\textbf{v}_{\text{rec}}(\textbf{x}_i)\exp\Big(-\frac{(\textbf{x}_j - \textbf{x}_i)^2}{2R_j^2}\Big)}{\sum_{j=1}^{N'}\exp\Big(-\frac{(\textbf{x}_j - \textbf{x}_i)^2}{2R_j^2}\Big)}.
\end{equation}
Then the predicted radial peculiar velocities are
\begin{equation}
    v_{\text{model},i} = \textbf{v}_{\text{smo}}(\textbf{x}_i) \cdot \textbf{x}_i.
\end{equation}

\begin{figure}
\centering \label{fig2}
    \includegraphics[width=0.7\linewidth]{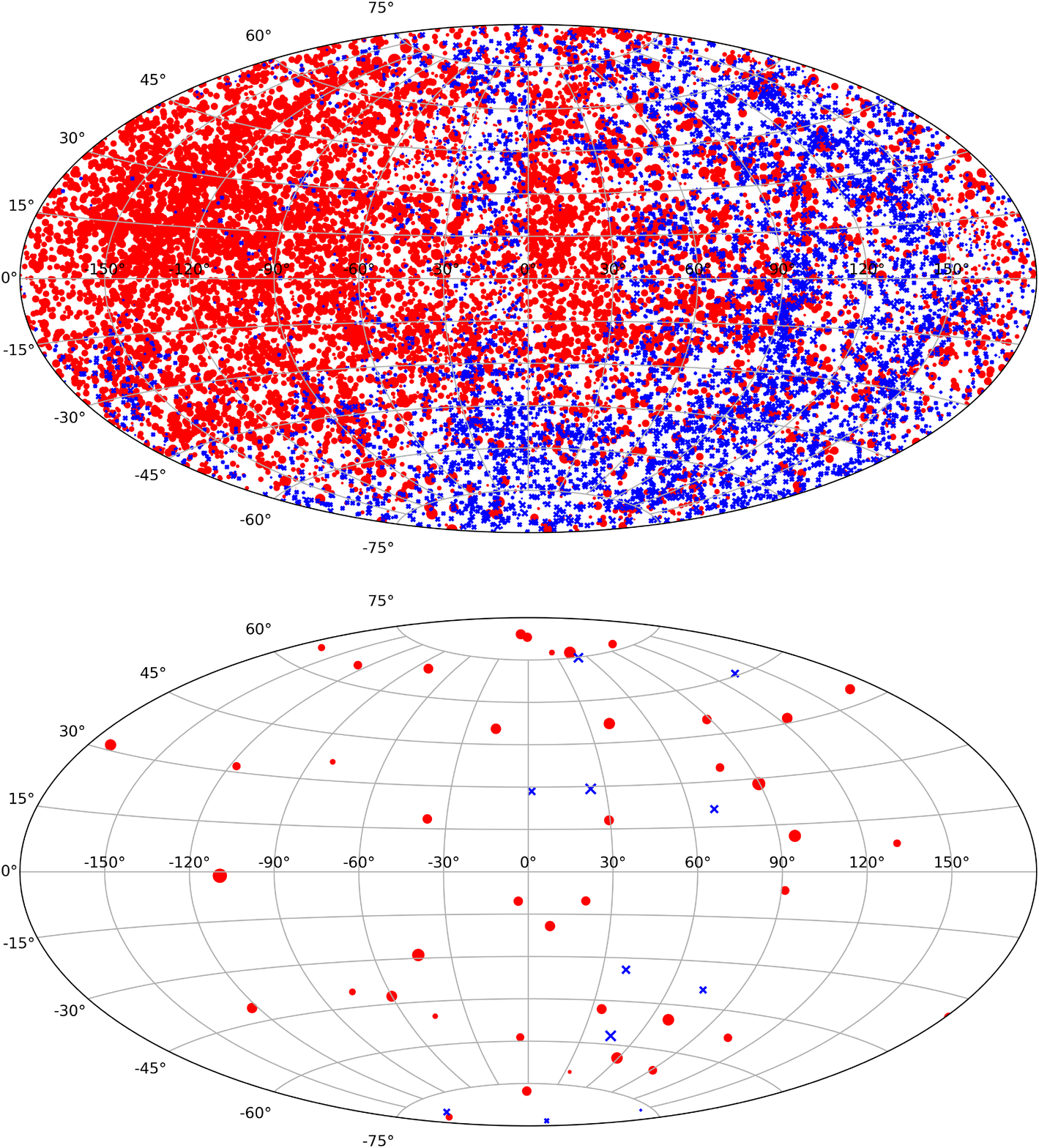}
    \caption{Peculiar velocities of PSC$z$ galaxies (15795 galaxies in total) and mock GW events projected in Galactic coordinates. Upper panel: the red circles and blue crosses represent PSC$z$ galaxies that are moving away from and moving towards us, respectively. Lower panel: the same as upper panel but for mock GW events. The size of markers is proportional to the magnitude of the line-of-sight peculiar velocity in each panel. The value of $H_0 = 67.8$ km s$^{-1}$ Mpc$^{-1}$ from Planck Collaboration is used.}
\end{figure}

\begin{figure}
    \includegraphics[width=0.45\textwidth]{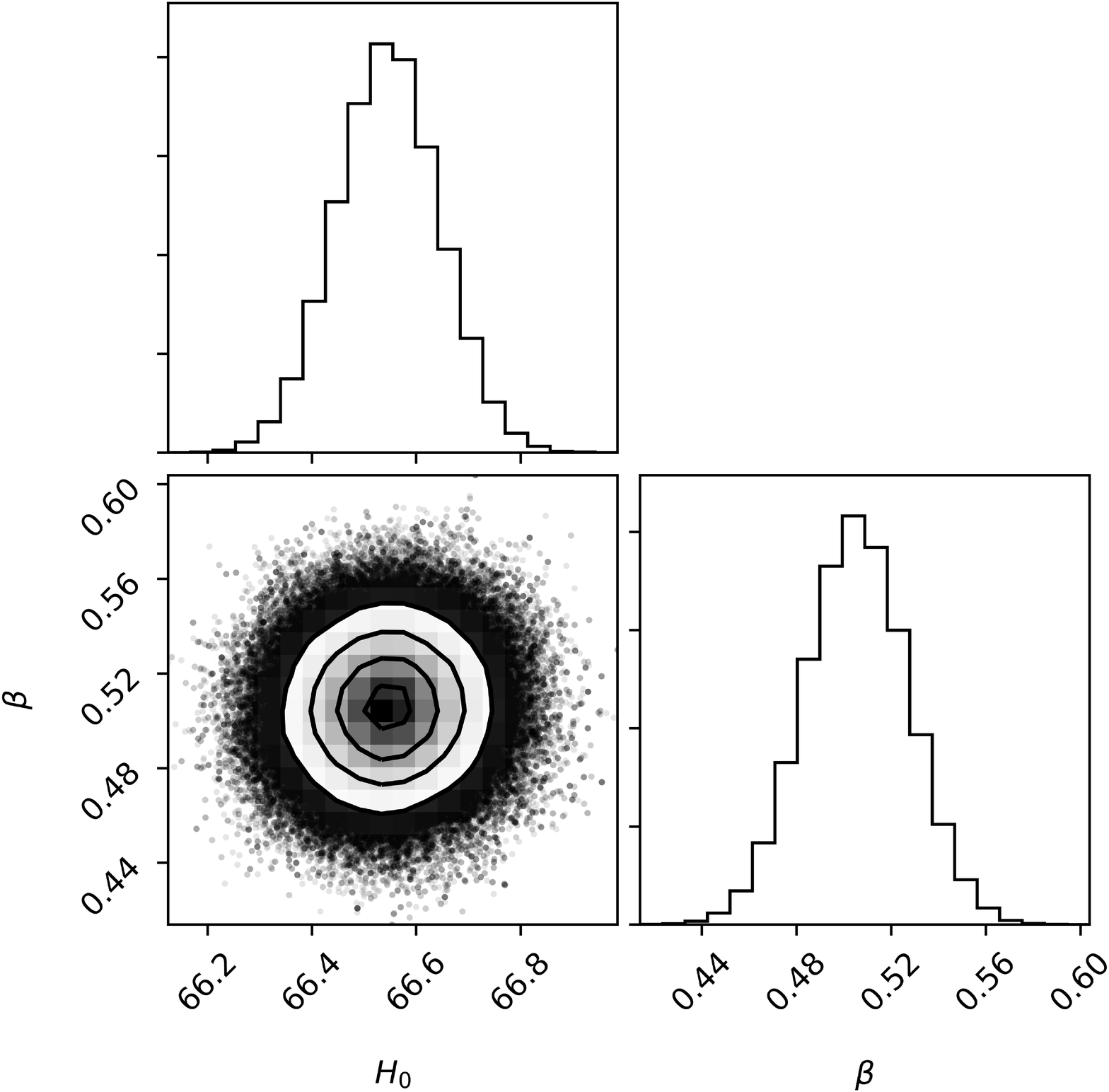}
    \caption{Constraints on $\beta$ and $H_0$ using $v-v$ comparison method from mock GW catalogue. Confidence contours (1$\sigma$ , 2$\sigma$ and 3$\sigma$) and marginalized likelihood distributions for the parameters $(H_0, \beta)$ are shown. The error of distance derived from GW waveform fitting is assumed as $1\%$. Our method can constrain the Hubble constant and $\beta$ simultaneously. }
\end{figure}

We construct a mock GW catalogue to compare measured peculiar
velocities with PSC$z$ survey\cite{Branchini99}. Abbott et al. (2016) estimated the
detection rate of BNS coalescences to be 4-80 per year for Advanced LIGO and Advanced Virgo after 2020, which will increase to 11-180 per year
after 2024 with more detectors\cite{Abbott16}. Fortunately, the ET
would observe $10^3-10^7$ BNS mergers per year\cite{Abernathy11}.
For the third-generation ground-based GW detectors (such as ET), the
precise localization and distance uncertainty from GW signals for
BNSs could be sufficient to directly identify the host
galaxies\cite{Zhao18} at $z<0.1$. Meanwhile, the proposed
space-based GW detector Big Bang Observer (BBO)'s angular resolution would be sufficient
to uniquely identify the host galaxy of compact binary
merger\cite{Cutler09}. Here we conservatively construct a mock GW catalog containing
90 events for the third-generation ground-based GW detector ET. The
spatial positions of BNSs are randomly sampled in the sky and their
volume density is uniform in the range [0, 0.045], which in
the farthest reach to 190 Mpc---the boundary of PSC$z$
reconstructed peculiar velocity field\cite{Ma12}. The simulated heliocentric redshifts of GW host galaxies are transformed to CMB-frame redshifts by
\begin{equation}
    1 + z_{\text{CMB}} = (1 + z_{\text{hel}})\bigg[1 + \frac{v_{\text{CMB}}}{c}(\hat{\textbf{n}}_{\text{CMB}}\cdot\hat{\textbf{n}})\bigg],
\end{equation}
where $\hat{\textbf{n}}$ is the direction cosine of GW source's sky
position, $\hat{\textbf{n}}_{\text{CMB}} = (263.99^\circ,
48.26^\circ)$ and $v_{\text{CMB}} = 369$ km/s \cite{Hinshaw09}. The
luminosity distance $d_L$ is obtained by fitting GW waveform signal.
In our simulation, they are generated from the redshift-distance
relation of Planck15/$\Lambda$CDM model\cite{Planck16} with normally
distributed discrepancy. In figure 2, we show the radial peculiar velocities of PSC$z$ galaxies\cite{Branchini99}
and our mock GW events projected in Galactic coordinates.

We employ python module \texttt{emcee}\cite{Foreman-Mackey13} to constrain $H_0$ and $\beta$ simultaneously.
The maximum likelihood estimation (MLE) is applied to the MCMC algorithm. The likelihood $L$ is the summation of many normal distributions
\begin{equation}
    L = \prod_{i=1}^{N}\frac{1}{\sqrt{2\pi}\sigma_i}\exp\bigg[\frac{-(v_{\text{pec,r},i} - \beta v_{\text{model},i})^2}{2\sigma^2_{v_{\text{pec,r},i}}}\bigg],
\end{equation}
where
\begin{equation}
	v_{\text{pec,r},i} = c z_i - H_0 d_{L,i}\,,
\end{equation}
and
\begin{equation}
    \bigg(\frac{\sigma_{v_{\text{pec,r}, i}}}{H_0 d_L}\bigg)^2 = \bigg(\frac{\sigma_{H_0}}{H_0}\bigg)^2 + \bigg(\frac{\sigma_{d_L}}{d_L}\bigg)^2.
\end{equation}
For the third-generation detectors, such as Einstein Telescope (ET)\cite{Abernathy11}, we adopt $\frac{\sigma_{H_0}}{H_0}\sim 1\%$ and $\frac{\sigma_{d_L}}{d_L}\sim 1\%$.
Then the log-likelihood is
\begin{equation}
    \ln L = -\frac{1}{2}\sum_{i=1}^{N}\bigg[\frac{-(v_{\text{pec,r},i} - \beta v_{\text{model},i})^2}{2\sigma^2_{v_{\text{pec,r},i}}} + \ln \big(2\pi \sigma^2_{v_{\text{pec,r}, i}}\big)].
\end{equation}
The priors of $H_0$ and $\beta$ are $[60, 80]$($\text{km}\, \text{s}^{-1} \,\text{Mpc}^{-1}$) and $[0, 1]$, respectively. Figure 3 shows the
$1\sigma$ to $3\sigma$ confidence contours and marginalized
likelihood distributions for $\{H_0, \beta\}$ from MCMC fitting. The constraints are
$H_0=66.55\pm0.10 ~\rm km~s^{-1} Mpc^{-1}$ (1$\sigma$) and
$\beta=0.504\pm0.022$ (1$\sigma$) respectively. The $\beta$ value is consistent with those of previous works, but with a smaller uncertainty.

The ability of our method depends on the measurement accuracies of the
Hubble constant $H_0$ and luminosity distance $d_L$. At present, the
value of $H_0$ can be determined at 1\% accuracy from CMB in the
$\Lambda$CDM model\cite{Planck16}. For BNSs, the distance accuracy
by Advanced LIGO can reach 10\%. The main uncertainty comes from
the errors of the distances measured by GW detectors. For the
third-generation detectors, such as ground-based ET\cite{Abernathy11}, space-based BBO\cite{Cutler09} and
Deci-Hertz Interferometer Gravitational wave Observatory (DECIGO)\cite{Kawamura06},
the distance uncertainties could be as low as 1\% \cite{Abernathy11}
and 0.1\% \cite{Cutler09,Kawamura06} in local universe,
respectively.  In the future, the uncertainty of the $H_0$ can also
be decreased to $0.1\%$ \cite{Cutler09,Liao17}, together with the
uncertainty of $d_L$ decreased to $0.1 \%$, the errors of peculiar velocities
can be determined to $\sim 10$ km/s including the uncertainty of
redshift determination at 100 Mpc. We also plot the errors of
peculiar velocities measured by the future GW detectors in figure 1.
From this figure, the future uncertainty of the peculiar velocity
can be about one order of magnitude smaller than the current
available uncertainties. On the other hand, once the inclination
angle of the GW is determined by other independent method, the
uncertainty of the distance can be reduced enormously, as the
distance and the inclination angle degenerate.

\begin{figure}
	\includegraphics[width=0.5\textwidth]{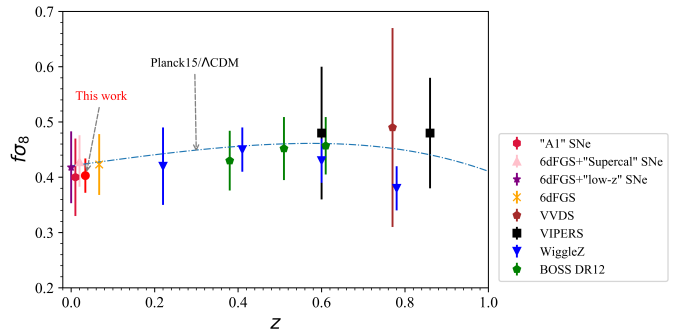}
	\caption{Estimates of $f\sigma_8(z)$ for different methods. Our
		constraint is shown as the red data point.  We also show past measurements of the $f\sigma_8(z)$ from 6dFGS at $z\simeq0$ and SNe\cite{Johnson14}, SNe\cite{Turnbull12}, 6dFGS and ``Supercal" SNe\cite{Huterer17}, 6dFGS $z=0.067$\cite{Beutler12}, WiggleZ\cite{Blake11}, VVDS\cite{Song09}, BOSS\cite{Satpathy17}, and VIPERS\cite{delaTorre17}. The growth parameter $f\sigma_8(z)$ as function of $z$ under Planck/$\Lambda$CDM is calculated with python module $\mathtt{CAMB}$\cite{Lewis00}, which is shown as dash-dotted line.}
\end{figure}

Next we will discuss the cosmological applications of the precise
peculiar velocities derived from our method. One well-known problem of standard $\Lambda$CDM model is the
tension between the relatively higher growth rate $f(z)\sigma_8(z)$
found in CMB experiments and the smaller one obtained from
large-scale galaxy surveys\cite{Macaulay13}. Recent study shows that
they are in 5$\sigma$ tension with each other\cite{Kazantzidis18}.
Using $\beta$ constrained by our method, we can measure the local
growth rate. Combining with $\sigma_{8,\text{g}}\simeq 0.80\pm0.05$
from PSC$z$ catalogue\cite{Hamilton02}, we have $f\sigma_8=
0.403\pm0.031$. Figure 4 displays the evolution of $f\sigma_8(z)$
with respect to redshift $z$ under Planck/$\Lambda$CDM
model\cite{Planck16} as well as the $f\sigma_8(z)$ data from various
large-scale structure surveys. We see
that our constraint with small error bar is very competitive with
the other existing constraints. On the other hand, the cosmic accelerated
expansion would be caused by the presence of a scalar field with an
evolving equation of state, or extensions of general relativity\cite{Dvali00,Carroll04}.
Although they produce similar expansion rates, different models
predict measurable differences in the growth rate of
structure\cite{Linder05,Huterer18}. Future GWs from BNSs could be observed by
ET\cite{Abernathy11} at a redshift of $z=2$. The growth rates at
different cosmic times can be measured by our method, which can
determine the origins of the cosmic accelerated expansion.

From redshift space distortion constraints, the current matter
density $\Omega_{m0}$ degenerates with $\sigma_{8,0}$\cite{Peacock01}. The
precise peculiar velocities obtained directly from GWs at low
redshifts can break this degeneracy. After correcting the
Alcock-Paczynski (AP) effect\cite{Alcock79} for the observed
$f\sigma_8(z)_{\text{meas}}$, the parameters $\Omega_{m0}$ and
$\sigma_{8,0}$ can be tightly constrained through
\begin{equation}
	\chi^2 = \sum_{i=1}\frac{[f\sigma_8(\text{meas})_i - f\sigma_8(\text{model})_i]^2}{\sigma_i^2}.
\end{equation}
The $1\sigma$, $2\sigma$
and $3\sigma$ confidence contours in the
$\Omega_{m0}$-$\sigma_{8,0}$ parameter plane are shown in figure 5,
where a flat $\Lambda$CDM model is assumed when calculating
$f\sigma_8(z)_{\text{model}}$. The
dashed contours show constraints from $f\sigma_8(z)$ at $z>0.2$ in
figure 5, while solid contours show the same constraints for the
above data plus the peculiar velocity measurements from our method
and other low-redshift $f\sigma_8$. We find that the best fit of $\Lambda$CDM model is
$\Omega_{m0} \simeq 0.33\pm 0.06$ and $\sigma_{8,0}\simeq 0.75\pm
0.05$. After adding the low-redshift peculiar velocity measurements, the
degeneracy can be broken.

Another serious problem called the $H_0$
tension\cite{Freedman17} says that the Hubble constant estimated from the local
distance ladder\cite{Riess16} is in $3.4\sigma$ tension with the
value fitted from Planck CMB data assuming $\Lambda$CDM model\cite{Planck16}. From our simulation, the precise peculiar
velocity measurements can constrain the Hubble constant with an uncertainty of 0.5\%, which can reconcile the $H_0$ tension.

\begin{figure}
\centering
    \includegraphics[width=0.40\textwidth]{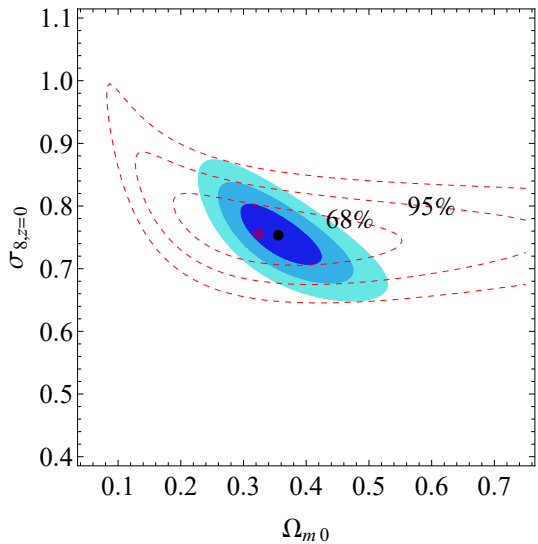}
    \caption{Confidence contours of $\Omega_{m0}$ and $\sigma_{8,0}$ in $\Lambda$CDM model. The red dashed contours show the $1\sigma$-$3\sigma$ confidence intervals using only high redshift ($z \ge 0.2$) $f\sigma_8$ measurements. The red point is the best fit. The blue contours show the same constraints for high redshift $f\sigma_8$ plus five low redshift data. The low-reshift $f\sigma_8$ can break the degeneracy between $\Omega_{m0}$ and $\sigma_{8,0}$. The best fit of $\Lambda$CDM model (black point) is $\Omega_{m0} \simeq 0.33\pm 0.06$ and $\sigma_{8,0}\simeq 0.75\pm 0.05$.}
\end{figure}

\section{Summary}\label{sec:summary}

The future GW detectors, such as ET\cite{ET}, LISA\cite{LISA}, DECIGO\cite{DECIGO} and BBO\cite{BBO}, will considerably enhance the angular resolution,
distance measurement and the detection rate of GW events. More GW events will be
detected and the host galaxies can be identified even without EM
counterparts. The peculiar velocities measured by our method would
be powerful cosmological tools. Although the
method we propose may be limited by the reconstructed peculiar velocity field, future galaxy surveys,
such as Euclid\cite{Euclid} and WFIRST\cite{WFIRST}, can
give high-quality peculiar velocity field in the near future.

\section*{Acknowledgements}
We thank the anonymous referee for useful comments and suggestions. We thank Enzo Branchini for providing the data of PSC$z$ peculiar velocity field and critical reading the manuscript.
This work is supported by the National Basic Research Program of China (973 Program, grant No.
2014CB845800), the National Natural Science Foundation of China
(grants U1831207, 11422325 and 11373022), and the Excellent Youth Foundation
of Jiangsu Province (BK20140016).

\end{document}